\documentclass[prb,preprint]{revtex4}

\usepackage{amsmath}
\usepackage{graphicx}

\begin{document}

\title{Newton's superb theorem: An elementary geometric proof}
\author{Christoph Schmid}
\email{chschmid@itp.phys.ethz.ch}
\affiliation{Eidgen\"ossische Technische Hochschule, 
Institute for Theoretical Physics, 8093 Zurich,
Switzerland}

\begin{abstract}
Newton's ``superb theorem'' for the gravitational $1/r^2$ force states 
that a spherically symmetric mass distribution
attracts a body outside as if the entire mass were concentrated
at the center.
This theorem is crucial for Newton's comparison of the Moon's orbit 
with terrestrial gravity (the fall of an apple),
which is evidence for the $1/r^2$ law.
Newton's geometric proof in the \textsl{Principia}
"must have left its readers in helpless wonder"
according to S. Chandrasekhar and J.E. Littlewood.
In this paper we give an elementary
geometric proof, which is much simpler 
than Newton's geometric proof and more elementary than proofs using calculus.
\end{abstract}

\maketitle

\section{Introduction}

Newton's ``superb theorem'' for the gravitational $1/r^2$ force
states that a spherically symmetric mass distribution attracts 
a body outside as if the entire mass were concentrated at the center.
The name ``superb theorem'' was used by S.\ Chandrasekhar
in \textsl{Newton's Principia for the Common Reader}\cite{Chandrasekhar}
and by J.\ W.\ L.\ Glaisher.\cite{Glaisher} 
See also I.\ B.\ Cohen and A.\ Whitman.\cite{Cohen.Whitman} 

The superb theorem is of crucial importance for Newton's comparison of 
terrestrial gravity (the fall of the apple)
with the orbit of the Moon. Newton wrote that\cite{apple.moon,Whiteside.moon} 
``The same year [1666] I began to think of gravity extending to the orb of the
Moon \ldots\ 
From Kepler's Rule of the periodical times of the Planets \ldots\ I deduced
that the forces which keep the Planets in their Orbs must [be] reciprocally
as the squares of their distances from the centers \ldots: and thereby
compared the force 
requisite to keep the Moon in her Orb with the force of
gravity at the surface of the earth, and found them to answer pretty nearly.''
The superb theorem is also crucial for an exact solution of the Kepler problem: 
Finding the orbit of a planet
under the attraction of the sun
(neglecting the attraction between planets)
for two spherically symmetric bodies 
with a $1/r^2$ force between mass elements. 

When Newton first made the comparison of the Moon's orbit with the fall of the
apple, he had no knowledge of the superb theorem.
He assumed that considering the mass of the Earth 
to be concentrated at the center 
to obtain the magnitude of the gravitational field on the surface of the Earth
was nothing more than an approximation:
``It might be \ldots\ accurate enough at greater distances \ldots\
wide off the truth near the surface of the planet \ldots, 
where the distances of the particles
are unequal \ldots.''\cite{wide.off.the.mark}
Newton suspected the superb theorem to be false until 1685,
as he wrote in his letter to Halley of 20 June 1686.\cite{Rouse.Ball}

Newton proved the superb theorem in 1685, one year before finishing the
\textsl{Principia}, 
in which the theorem was published as Proposition 71 of Book I in 1687. 
We know from Newton's own words 
that ``he had no expectation of so beautiful a result
till it emerged from his mathematical 
investigation.''\cite{so.beautiful.a.result}
In Chandrasekhar's words: ``The superb theorem is most emphatically
against common sense \ldots\ unless one had known 
its truth already.''\cite{against.common.sense}

Chandrasekhar translated Newton's geometrical proof into modern
notation.\cite{modern.notation} 
See also Refs.~\onlinecite{Weinstock}
and \onlinecite{Cushing.proof}.
Chandrasekhar reported the comment by Littlewood that
Newton's geometrical construction for the 
proof ``must have left its readers in 
helpless wonder.''\cite{helpless.wonder, Littlewood.helpless.wonder}
D.T. Whiteside characterized Newton's proof as
``opaque and overlong.''\cite{Whiteside.opaque.overlong}

In this paper we give a geometric proof of the superb theorem, 
which is much simpler than Newton's geometric proof.
Our geometric proof has four elementary steps and is 
also simpler than the proofs using calculus in 
textbooks (see, for example, Refs.~\onlinecite{Feynman} 
and \onlinecite{Marion}).
Our proof is suitable for introductory physics courses without calculus
and is accessible to good high school students.
In typical graduate textbooks on classical mechanics the superb theorem 
is not mentioned.\cite{grad.texts}

In Sec.~\ref{section.70} we give Newton's elementary geometric proof 
for a test mass {\it inside} a spherical mass-shell. 
The geometry of this proof is closely related to the geometry 
of our new proof in Sec.~\ref{proof.superb.theorem} for 
Newton's superb theorem, that is, for a test mass {\it outside} 
a spherical mass-shell. 
In Sec.~\ref{infinitesimals} we discuss the role of infinitesimals 
in the method of Newton's geometric
proofs, a method that is a reformulation of ``the manner of ancient
geometers,'' in particular of Archimedes. In Sec.~\ref{sec.equiv.theorem} we
convert the problem of the superb theorem to an equivalent problem with the
geometry of {\it one point inside a spherical shell}:  
calculating the spherical average of the radial accelerations at all
observation points at a distance $r$ from the center due to only one source
point. In Sec.~\ref{solution.equiv.problem} we give the elementary geometric
solution of this equivalent problem and hence the proof of Newton's superb
theorem. The geometry of the solution of the equivalent problem is much
simpler than the geometry of Newton's derivation. 
The solution of the equivalent problem is a geometric
derivation of Gauss's law in integral form for a closed spherical surface 
without using vector calculus. 
In Sec.~\ref{sec.comments} we give two more
propositions of Newton based on the superb theorem
(relevant for the falling apple and for the system Moon-Earth), 
comment on Littlewood's proof of the superb theorem, 
and note that the solution of the equivalent problem, 
Gauss's law in integral form for Newtonian gravity, 
is the same as the $R^{\,0}_{~0}$ Ricci-tensor equation in integral form 
of Einstein's gravity (general relativity) 
for sources that are nonrelativistic relative to the observer.

\section{Newton's geometric proof for a test mass 
within a spherical mass shell}
\label{section.70}

For a test mass inside
a homogeneous spherical shell of matter,
Newton's Proposition~70 of Book I states:
``If to every point of a spherical surface there tend equal centripetal forces
decreasing as the square of the distances from those points, 
I say, that a corpuscule placed within that surface
will not be attracted by those forces any way.''\cite{Prop.70}

For the analogous case of the $1/r^2$ law of electrostatics, Newton's
proposition 70 was tested by the null experiment 
of Henry Cavendish in 1773.\cite{Cavendish}

Newton's elementary geometric proof of Proposition~70
(Ref.~19)
is shown in Fig.~1: With the test body at the observation point $P$ 
as the apex, construct any cone of infinitesimally 
small solid angle $d \Omega$
that intersects the spherical source-shell in both directions 
by the source-surface areas
$dS$ and $dS'$ around the source points $Q$ and $Q'$.
The attractions per unit test mass at $P$ by these two areas, 
$dS$ and $dS'$, are equal and opposite
because (1) the gravitational force decreases as the square of the distance,
while the surface areas of the source, cut out by the cone 
of given solid angle $d \Omega$,
increase as the square of the distance;
and
(2)~the angles $\theta$ between the infinitesimally 
thin cone (in both directions) and the normals to the source sphere 
at both intersection points
(radial lines $OQ$ and $OQ'$)
are equal.
Because the entire solid angle around the point $P$
can be divided into such double cones, the resultant attraction is zero. Q.E.D.

\section{Infinitesimals in Newton's geometric proofs}
\label{infinitesimals}

Newton's geometric proof of Proposition~70 (our Sec.~II) 
uses his ``evanescent quantities," that is, 
vanishing quantities (infinitesimals), 
and in particular his ``method of ultimate ratios''
(title of Sec.~I of Book~I):\cite{Cohen.Whitman.ultimate.ratios}
In the Scholium at the end of this section 
Newton explains ``~\ldots\ the ultimate ratio of evanescent 
quantities is to be understood not as the ratio of quantities 
before they vanish \ldots, but the ratio with which they vanish.'' 
Newton emphasizes that
``ultimate ratios with which quantities vanish 
are not ratios of ultimate quantities, but limits with which the 
ratios of quantities decreasing without limit are continually approaching,
and which they can approach so closely that their difference 
is less than any given quantity.''\cite{details.ultimate.ratio}
In the case of the proof of Prop.~70, 
it is the ultimate ratio of the magnitudes of the attraction
by the evanescent (that is, infinitesimal) source areas.
Newton does not write $(dS, dS'),$ instead he discusses 
the analogous arcs in his plane figure and in his proof.
In his notation he uses the endpoints of the small arcs.
But Newton emphasizes that his evanescent arcs can be replaced by straight lines:
Newton's evanescent arcs are infinitesimal.

Throughout the Principia, Newton uses ultimate ratios, 
with which he is ``solidly in a contemporary tradition of 
Fermat, Blaise Pascal, Huygens, James Gregory, 
and Isaac Barrow.''\cite{contemporary.tradition}
In this scholium Newton wrote that ``I have presented these 
lemmas \ldots\ to avoid the tedium of \ldots\ lengthy proofs 
by `reductio ad absurdum' in the manner of the 
ancient geometers.''\cite{ancient.geometers}

Euclid and Archimedes presented early forms of infinitesimal thinking
in the ``method of exhaustion'' (of Eudoxus) combined with ``reductio ad absurdum.''
\cite{Katz.Euclid.Archimedes,Archimedes.infinitesimals,Wussing} 
An example is the proof by Archimedes that the area inside a circle 
is equal to the product of the radius times half of the circumference:
In the method of exhaustion
one can cut the circular disc radially into an increasing number $N$ 
of identical slices,
one compares the thin slices with inscribed (respectively, circumscribed) triangles;
that is, one compares the circular disc with an inscribed 
(respectively, circumscribed) $N$-sided polygon.
$N$ can be chosen arbitrarily large. 
In a simplification by Leonardo da Vinci one considers $N$ 
even and reassembles the triangles in an alternating sequence
to obtain 
an area arbitrarily close to a rectangle for $N$ arbitrarily large.
If somebody claimed that the area of the circular disc is some given number 
which is smaller than the radius times half the circumference,
one ``reduces this to absurdum'' (proof by contradiction) 
by choosing $N$ so large that the total area of the inscribed triangles 
is larger than the claimed result. 
One then repeats the argument with circumscribed triangles.

Newton's geometric proofs 
in the Principia use evanescent quantities (i.e. infinitesimals). 
But these proofs do not use the machinery of calculus developed by Newton and Leibniz.

\section{\label{proof.superb.theorem}The elementary geometric 
proof of the superb theorem}

In Sec.~\ref{firstsubsec} we convert the problem of the superb theorem
(Proposition~71, a test mass outside a homogeneous spherical shell)
to an equivalent problem using spherical symmetry. In Sec.~\ref{solution.equiv.problem}
we give the elementary geometric solution of the equivalent problem
using a method similar to the method used by Newton for Proposition~70
(our Sec.~II).

\subsection{\label{firstsubsec}Spherical symmetry: 
Translation to an equivalent problem }
\label{sec.equiv.theorem}

The superb theorem considers a spherical shell of matter as the source of gravity and 
gives the (radial) acceleration of a test particle
at the observation point $P$ outside the source shell
(observation point at a distance $r$ from the center), see Fig.~2(a).
Instead we analyze the equivalent problem
of one source particle located at $Q$, as shown in Fig.~2(b)
and find the radial components of the accelerations averaged 
over a spherical observation surface outside the source particle
(observation surface at a distance $r$ from the center),
as shown in Fig.~2(b). 
The steps needed for calculating this spherical average 
are explained before Eqs.~(3) and (4).

The proof of the equivalence of the two problems is given in two steps:

\begin{enumerate}
\item We consider the average over the spherical observation surface of the
  radial component of the acceleration, $g_{\rm radial} = {\bf g} \cdot {\bf
    r}/r$, measured outside the spherical mass shell at the distance $r$ from
  the center. This average is obtained by multiplying $g_{\rm radial}$ by the
  area element $dS$ on the observation surface, summing the contributions, and
  then dividing by the total surface area. The average is equal to the radial
  acceleration at any one of the observation points, such as the point $P$,
  because of the spherical symmetry of the source 
  in the first problem, as shown in Fig.~2(a). 
  This first step appears trivial, but it is necessary to make the second step possible.

\item The spherical average on the observation sphere 
of the radial component of the acceleration yields the same contribution 
from each single source element of equal mass in the spherical source shell, 
such as $Q$, because of the spherical symmetry of the 
observation surface [see Fig.~2(b)]. 
Hence, it is sufficient to consider the contribution of 
only one source mass-element at the point $Q$, that is, 
the equivalent problem of Fig.~2(b), and later
sum over the equal contributions of all other source elements of equal mass.

\end{enumerate}
The crucial point is that the geometry is much simpler 
for the solution of our equivalent problem, Fig.~2(b) (one point inside a shell), 
than the complicated geometry needed by Newton to solve the original problem, 
Fig.~2(a) (one point outside a shell).

These steps complete the proof of the equivalence of the two problems. 
The equivalent problem and its solution in Sec.~\ref{solution.equiv.problem} 
is Gauss's law in integral form for a spherical surface.

\subsection{\label{solution.equiv.problem}Geometric solution 
of the equivalent problem}

The equivalent problem of Fig.~2(b)
can be solved with the elementary geometric method 
shown in Fig.~3; the method is analogous to Newtons's method to prove 
proposition~70 (given in Sec.~\ref{section.70}).
We use the source point $Q$ as the apex and consider an arbitrary ray
which starts at $Q$,
goes in one direction,
and hits the spherical surface at the observation point~$P$.
At $P$, the magnitude of the gravitational acceleration $g$ is $g_P = G m/L_{QP}^2$,
where $m$ is the small (infinitesimal) source mass at $Q$,
and $G$ is Newton's constant. The angle $\theta$ is
the angle between the ray $QP$ and the radial direction $OP$.
The radial component of the gravitational acceleration at $P$ is
\begin{equation}
g_{\rm radial} (P) = - (G m/L^2_{QP})\cos \theta.
\label{radial.component}
\end{equation}
Around this ray from $Q$ to $P$, we consider a cone starting at the apex $Q$ and
of infinitesimally small solid angle $d \Omega$,
which will intersect the spherical observation surface 
by the surface area $dS$ given by
\begin{equation}
dS = (L_{QP}^2) d \Omega / \cos \theta.
\label{surface.area}
\end{equation}
The first step in calculating the average of the radial accelerations 
over the observation sphere
is to form the product
of the radial acceleration given by Eq.~(\ref{radial.component}),
with the weight (for averaging), which is the surface area element $dS$
(intersected by the cone) 
given by Eq.~(\ref{surface.area}):
\begin{equation}
g_{\rm radial} \, dS = - G m \, d \Omega.
\label{product}
\end{equation}
Note the two cancellations in $(g_{\rm radial} \, dS)$, which are the basis
for the simplicity of our proof:
\begin{enumerate}
\item The inverse square of the distance in the force law, $1/L^2_{QP}$,
  cancels $L^2_{QP}$ 
in the surface area $dS$ for a cone with solid angle $d \Omega$. 

\item The ratio $\cos \theta$ of the radial to the total acceleration cancels
  $1/\cos \theta$ 
in the surface area $dS$ for a cone with solid angle $d \Omega$.
\end{enumerate}
The second step in calculating the spherical average is 
to sum $(g_{\rm radial} \, dS)$ in Eq.~\eqref{product}
over all surface area elements $dS$, that is, over the entire observation surface.
The sum over $d\Omega$ is $4 \pi$.
The third step in taking the spherical average of $g_{\rm radial}$
is to divide by the total weight for averageing,  
which is the total area of the spherical observation surface, $4 \pi r^2$.
We thus obtain
\begin{equation}
\langle g_{\rm radial}\rangle^{\rm spherical\, average} = - G m/r^2,
\label{average.radial.acceleration}
\end{equation} 
This result for the spherical average of the radial accelerations is the same 
as if the single point mass $m$ were placed at the origin.

The conversion back to the original problem, Newton's superb theorem, now follows.
For a spherically symmetric source shell,
the acceleration is radial, and because it is the same all over the observation sphere,
there is no need to take the average over the spherical observation shell.
Therefore $\vec{g} = - (\vec{r}/r) G M_{\rm tot}/r^2$ at every observation point 
outside the source shell.
This result completes our geometric proof of Newton's superb theorem.

Equation~(\ref{average.radial.acceleration}), which is equivalent to Newton's
superb theorem, 
is Gauss's law in integral form\cite{Jackson,Feynman.Gauss} for a spherical surface. 
Gauss's law for the flux integral over a closed surface $S$ enclosing the volume $V$, 
$\oint_{S} {\bf g} \cdot {\bf dS} = - 4 \pi G M_{V}$, a concept from vector calculus, 
is not taught in undergraduate or high school courses on Newtonian mechanics. 
The superb theorem follows simply from 
Equation~(\ref{average.radial.acceleration}), Gauss's law in integral form 
for a closed spherical surface.

In taking the average of $g_{\rm radial}$ over the observation sphere, 
we had to sum over all infinitesimal surface area elements $dS$.
This sum is called an ``ultimate sum'' in the Principia,\cite{ultimate.sum} which 
corresponds to an integration in today's language. 
In the integration over $d\Omega$, the integrand is a constant, as given in
Eq.~\eqref{product}.
Therefore the integration is reduced to calculating the surface of the sphere.

\section{\label{sec.comments}Comments}

We have given a simple geometric proof of the superb theorem
along the lines of Newton's simple geometric proof of Proposition 70.

For a source with spherical symmetry and a radius-dependent density 
we add up the accelerations outside the source caused by all shells, 
and conclude that the result
is the same as if all the mass
was placed at the origin: Newton's Proposition 74, which is needed for
the falling apple.

For two dissimilar spheres with different spherical radius-dependent 
density distributions
we add up the contributions of matter shells and conclude that (in Newton's
words 
at the end of Proposition 76) 
``the whole force with which one of these spheres attracts the other
will be inversely proportional to the square of the distance of the centers,''
which Newton found ``very remarkable.'' This proposition is needed for
the attraction between Moon and Earth.

According to Chandrasekhar,\cite{Littlewood.integration} ``J.\ E.\ Littlewood
has conjectured 
(though I do not share in the conjecture \ldots)
that Newton had perhaps first constructed a proof
based on {\it calculus} which `we can infer with some possibility what the
proof was'.'' 
Chandrasekhar gives Littlewood's conjectured proof,
which consists of 14 equations.
Our geometric proof with four steps is simpler than Littlewood's proof using calculus.

Equation~(\ref{average.radial.acceleration}) for Newtonian gravity
is the same as 
the $R^0_{\, \, 0}$ Ricci-tensor equation in integrated form 
for Einstein gravity (general relativity)
for sources with $v \ll c$ relative to the observer.
This Einstein equation determines the spherical average 
of the radial geodesic deviation
(that is, of the radial accelerations of free-falling test particles 
relative to a free-falling observer).


\section*{Figures}


\begin{figure}[h]
\begin{center}
  \includegraphics{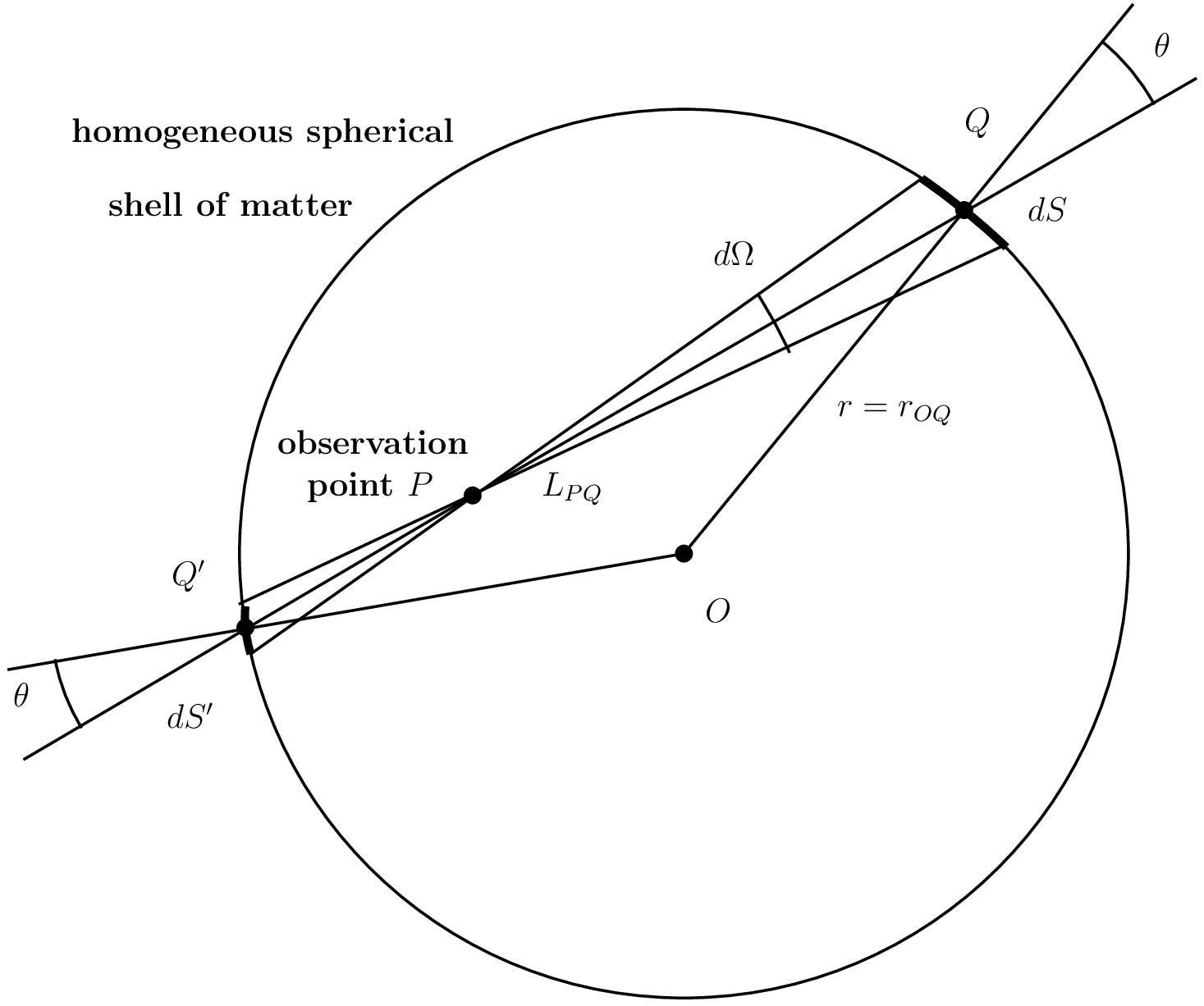}
   \caption{\label{fig:1} 
Newton's proof for a test mass inside a homogeneous spherical shell of matter
(Proposition~70). O is the center of a homogeneous spherical shell of matter,
P is the position of a test particle, 
$d\Omega$ is the solid angle of an infinitesimally thin cone with $P$ as the
apex, 
$dS$, $dS'$ are intersection areas (around $Q$ and $Q'$) of the two sides 
of the cone with the spherical shell of matter, and $\theta$ equals 
the two equal angles between $PQ$ and$PQ'$ 
to the normals on the matter shell, $OQ$ and $OQ'$.}
\end{center}
\end{figure}


\begin{figure}[h]
\begin{center}
  \includegraphics{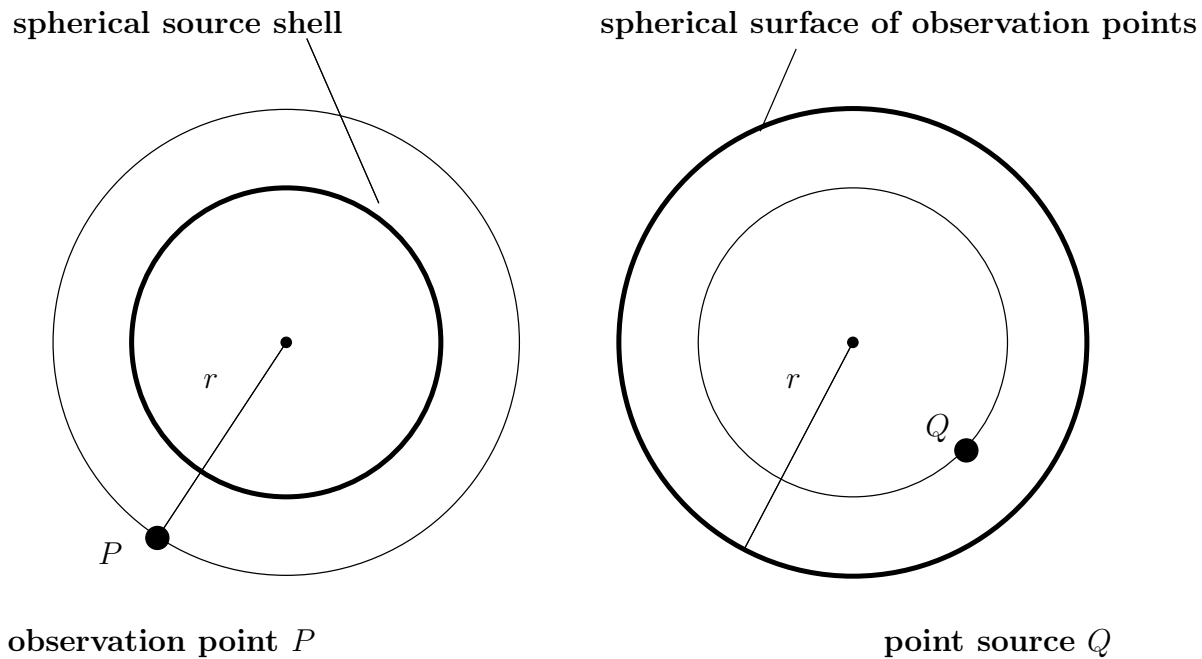}
   \caption{\label{fig:2}
The two equivalent problems. (a) Spherical source shell (solid inner circle)
and test point mass at observation point $P$ {outside} the source-shell. 
(b) Spherical observation surface (solid outer circle) 
and one {source point mass} at $Q$
inside the observation surface: 
Our proof for the geometry with one point 
inside a spherical surface, Fig.~2(b), 
is much simpler than Newton's proof 
for the geometry with one point outside a spherical surface, Fig.~2(a).}
\end{center}
\end{figure}


\begin{figure}[h]
\begin{center}
 \includegraphics{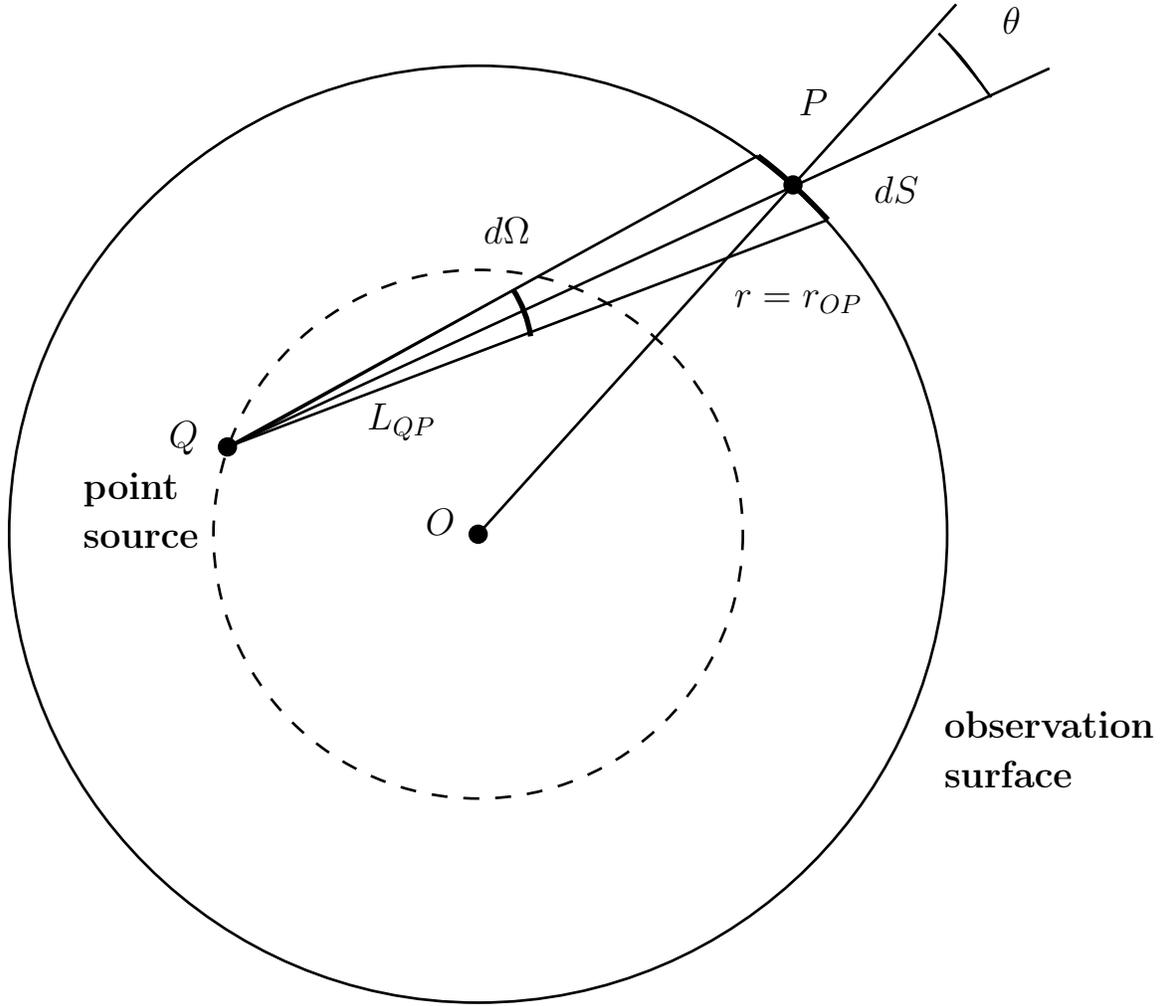}
\caption{\label{fig:3}
Elementary proof of Newton's superb theorem (Proposition~71). 
$Q$ is the source point, $P$ is the observation point, 
$O$ is the center of the observation surface, 
$\theta$ is the angle between $QP$ and the normal $OP$ 
on the observation surface, 
$d\Omega$ is the solid angle of the narrow cone with $Q$ as the vertex, 
and $dS$ is the surface area of the intersection of the cone 
with the observation surface.}
\end{center}
\end{figure}

\end{document}